\documentclass[pre,twocolumn,floatfix,superscriptaddress,longbibliography,nofootinbib]{revtex4-1}
\RequirePackage[sort&compress]{natbib}
\usepackage{natbib}	

\usepackage{multirow}

\usepackage{amssymb,amsfonts,amsmath}
\usepackage{bm}
\usepackage[pdftex]{graphicx}
\usepackage{xcolor}
\usepackage{setspace}
\usepackage{subfigure}

\definecolor{darkGray}{RGB}{153,153,153}
\definecolor{darkBlue}{RGB}{37,113,161}
\definecolor{darkGreen}{RGB}{113,161,37}
\definecolor{darkRed}{RGB}{186,21,24}

\usepackage{booktabs}
\newcommand{\mytoprule}{\specialrule{0.1em}{0em}{0em}}
\newcommand{\mybottomrule}{\specialrule{0.1em}{0em}{0em}}

\usepackage{textcomp}

\usepackage[activate={true,nocompatibility},final,tracking=true,kerning=true,spacing=true,factor=1100,stretch=10,shrink=10]{microtype}
\microtypecontext{spacing=nonfrench}

\usepackage{hyperref}
\usepackage{cleveref}
\crefname{equation}{Eq.}{Eqs.}
\crefname{figure}{Fig.}{Figs.}
\crefname{table}{Table}{Tables}

\begin{document}

\newcommand{\enter}{\curvearrowleft}
\newcommand{\exit}{\curvearrowright}
\newcommand{\intra}{\circlearrowright}

\title{Efficient community detection of network flows\\ for varying Markov times and bipartite networks}

\author{Masoumeh Kheirkhahzadeh}
\affiliation{Department of IT and Computer Engineering, Iran University of Science and Technology, Teheran, Iran}
\affiliation{Integrated Science Lab, Department of Physics, Ume{\aa} University, SE-901 87 Ume{\aa}, Sweden}

\author{Andrea Lancichinetti}
\affiliation{Integrated Science Lab, Department of Physics, Ume{\aa} University, SE-901 87 Ume{\aa}, Sweden}

\author{Martin Rosvall}
\email{martin.rosvall@umu.se}
\affiliation{Integrated Science Lab, Department of Physics, Ume{\aa} University, SE-901 87 Ume{\aa}, Sweden}

\begin{abstract}
Community detection of network flows conventionally assumes one-step dynamics on the links. For sparse networks and interest in large-scale structures, longer timescales may be more appropriate. Oppositely, for large networks and interest in small-scale structures, shorter timescales may be better. However, current methods for analyzing networks at different timescales require expensive and often infeasible network reconstructions. To overcome this problem, we introduce a method that takes advantage of the inner-workings of the map equation and evades the reconstruction step. This makes it possible to efficiently analyze large networks at different Markov times with no extra overhead cost. The method also evades the costly unipartite projection for identifying flow modules in bipartite networks. 
\end{abstract}

\maketitle

\section*{Introduction}

Researchers often represent interactions between components in social and biological systems with networks
of nodes and links, and use community-detection algorithms to better understand their large-scale structure.
Depending on the system under study and the particular research question, the scale of interest varies.
For an initial investigation, a bird's-eye-view of the entire system may be most appropriate, while 
a more detailed study most likely will require a finer scale. Methods for 
extracting hierarchically nested modules at different scales do exist \cite{rosvall2011multilevel,peixoto2014hierarchical}, but there may still be a need for identifying large-scale structures at specific scales \cite{delvenne2010stability,schaub2012markov}.

When the links represent network flows, modeling the dynamics at different Markov times is a natural way to capture the large-scale structures at different scales \cite{Schaub:2012ct}. In this approach, the original network is rebuilt such that one flow step along a link of the rebuilt network corresponds to the desired number of flow steps on the original network. However, this approach is inefficient for large networks, because the rebuilt network can be dense to the degree that storage and further analysis is infeasible. To overcome this problem, we introduce an efficient method that operates directly on the original network. The method takes advantage of the mechanics of the information-theoretic community-detection method known as the map equation \cite{rosvall2008maps} with no extra overhead cost.

Integrating the Markov time scaling with the map equation also allows for efficient community detection of network flows in bipartite networks. Most approaches for bipartite networks build on configuration models, in particular modularity \cite{Barber:2007kp,guimera2007module,crampes2014unified}, or stochastic block models \cite{peixoto2013parsimonious,larremore2014efficiently}. An alternative is to project the bipartite network into a unipartite network and perform the analysis on the unipartite network. For most assortative networks, such a projection does not destroy any valuable information \cite{everett2013dual}. However, the projection can give an overload of links and be infeasible for large networks. Therefore, the analysis of network flows derived from bipartite networks, such as unipartite collaboration networks obtained from projections of author-paper bipartite networks \cite{alzahrani2014community}, can greatly benefit from evading the projection into overly dense networks. With the map equation for varying Markov times, we can achieve this because a bipartate to unipartite projection corresponds to doubling the Markov times.

We begin by explaining the generalization of the Map equation to different Markov times and then introduce the bipartite generalization.

\section*{Network flow modules at different Markov times}

The map equation measures how well a partition of nodes in possibly nested and overlapping modules can compress a description of flows on a network. Because compression is dual to finding regularities in the data \cite{shannon1948mathematical}, the modules that gives the best compression also are best at capturing the regularities in the network flows. The network flows can be explicit flow data, such as the number of passengers traveling between cities, or be modeled by a random walker guided by the constraints set by a directed, weighted network, such as information flows on a citation network.

In the standard formulation of the map equation, a random walker is modeled as a \emph{discrete-time Markov process} and its position in the network is encoded at every transition. In this way, the transition rate of a random walker as well as the encoding rate is 1. Specifically, the discrete-time transition matrix associated with the network, $T_\mathrm{D}$, labeled with subscript D for \emph{discrete}, induces flows between nodes visited with probability $\vec{p}$ by the discrete-time Markov process
\begin{align}\label{eq:discretemarkov}
\vec{p}_{k+1} = \vec{p}_kT_{\mathrm{D}}.
\end{align}
\clearpage
Schaub \textit{et al.\ } generalized the map equation to different Markov times by using the corresponding \emph{continuous-time Markov process},
\begin{align}\label{eq:continuousmarkov}
\vec{\dot{p}} = -\vec{p}\left(I-T_\mathrm{D}\right),
\end{align}
with $I$ for the identity matrix \cite{Schaub:2012ct}. The continuous-time Markov process has exponentially distributed holding times at each node that correspond to Poisson-distributed transitions at average rate 1 \cite{delvenne2010stability,lambiotte2014random}. With uniform time steps $t$, the continuous-time Markov process is therefore equivalent to the discrete-time process 
\begin{align}
\vec{p}_{k+1} = \vec{p}_kT_\mathrm{C}(t),
\end{align}
with the continuous-time transition matrix
\begin{align}\label{eq:continuoustransition}
T_\mathrm{C}(t) = e^{-t(I-T_\mathrm{D})} = \sum_{i=0}^{\infty}\frac{t^ie^{-t}}{i!}T_\mathrm{D}^i,
\end{align}
labeled with subscript C for \emph{continuous}. By using this transition matrix, Schaub \textit{et al.\ } showed the effects of shorter and longer Markov times $t$ between encodings \cite{Schaub:2012ct}. Shorter Markov times than 1 mean that the average transition rate of a random walker is lower than the encoding rate of its position, such that the same node will be encoded multiple times in a row. As a result, the map equation will favor more and smaller modules. Oppositely, longer Markov times mean that the average transition rate is higher than the encoding rate, such that not every node on the trajectory will be encoded, and the map equation will favor fewer and larger modules. When a two-level solution is preferred over hierarchically nested modules of different sizes, changing the Markov time can in this way highlight salient flow modules at specific scales \cite{Schaub:2012ct}.

\subsection*{The map equation for varying Markov times}

In detail, for a given partition of nodes into modules, the original map equation for a discrete process at Markov time 1 measures the per-step minimum modular description length of flows on the network. For unique decoding of the flow trajectory from one step to another, the modular coding scheme is designed to only require memory of the previously visited module and not the previously visited node. The map equation therefore has one or, for hierarchically nested modules, more \emph{index codebooks} for encoding steps between modules and \emph{modular codebooks} for encoding steps within modules. Minimizing the map equation over all possible network partitions therefore gives the assignments of nodes into modules that best capture modular flows on the network. That is, the map equation can identify modules in which flows stay for a relatively long time. 

As input, the map equation takes the ergodic node visit-rates $p_\alpha$, module exit-rates $q_{i\exit}$, and module enter-rates $q_{i\enter}$ of the flow trajectory for nodes $\alpha=1 \ldots n$ and modules $i=1 \ldots m$. It estimates the average code length of each codebook from the Shannon entropy, which sets the theoretical lower limit according to Shannon's source code theorem \cite{shannon1948mathematical}. With $p_{i\intra}=q_{i\exit}+\sum_{\alpha \in i}p_\alpha$ for the total rate of use of module codebook $i$, the per-step average code length of events $\mathcal{P}^{i}$ in module $i$ is
\begin{align}\label{eq:modulecodelength}
H(\mathcal{P}^{i}) = - \frac{q_{i\exit}}{p_{i\intra}} \log\frac{q_{i\exit}}{p_{i\intra}} 
 -\sum_{\alpha \in i} \frac{p_{\alpha}}{p_{i\intra}}\log\frac{p_{\alpha}}{p_{i\intra}}.
\end{align}
Similarly, with $q_{\enter} = \sum_{i=1}^{m} q_{i\enter}$ for the total rate of use of the index codebook in a two-level description, the per-step average code length of module enter-events $\mathcal{Q}$ is
\begin{align}\label{eq:indexcodelength}
H(\mathcal{Q}) = - \sum_{i=1}^{m}\frac{q_{i\enter}}{q_{\enter}} \log\frac{q_{i\enter}}{q_{\enter}}.
\end{align}
With modular map $\mathsf{M}$ and the rate of use of each codebook taken into account, the map equation takes the form
\begin{align}\label{eq:mapeqmarkov1}
L(\mathsf{M}) = q_{\enter} H(\mathcal{Q}) + \sum_{i=1}^{m}p_{i\intra}H(\mathcal{P}_i). 
\end{align}

For an efficient generalization of the map equation to Markov times other than 1, we first linearize in $t$ and $T_\mathrm{D}$ the continuous-time transition matrix $T_\mathrm{C}(t)$ in \cref{eq:continuoustransition}. For $t < 1$, $(1-t)I + tT_\mathrm{D}$ is a valid approximation, but we are also interested in Markov times greater than 1. Thus, we consider the linearized transition matrix
\begin{align}\label{eq:linearizedtransition}
\tilde{T}_\mathrm{C}(t) = \begin{cases}
               (1-t)I + tT_\mathrm{D}     & t < 1\\
               tT_\mathrm{D}              & t \ge 1,
           \end{cases}
\end{align}
which captures Markov times below 1 with self-links and Markov times above 1 with transition rates proportional to the average rate of the underlying Poisson process. Moreover, at Markov time 1 it recovers the discrete-time transition matrix in \cref{eq:discretemarkov}. 

This linearization also has an appealingly simple effect on the map equation. For Markov time $t$, all node visit rates $p_\alpha$ remain the same, since the relative visit rates at steady state do not depend on how often the visits are sampled. However, the module exit-rates $q_{i\exit}$ and module enter-rates $q_{i\enter}$ change linearly with the Markov time, since the number of random walkers that moves along any link between nodes during time $t$ is directly proportional to $t$ as shown in \cref{eq:linearizedtransition}. Therefore, 
\begin{align}
q_{i\exit}\to t q_{i\exit} &\equiv q_{i\exit}(t)\label{eq:markovexit}\\
q_{i\enter}\to t q_{i\enter} &\equiv q_{i\enter}(t)\label{eq:markoventer}.
\end{align}
The rescaled module exit- and enter-rates affect both the module code length in \cref{eq:modulecodelength} and the rate of use of all codebooks. With $(t)$ for the Markov time, the map equation for Markov time $t$ takes the form
\begin{align}\label{eq:mapeqmarkovt}
L(\mathsf{M},t) = q_{\enter}(t) H(\mathcal{Q}) + \sum_{i=1}^{m}p_{i\intra}(t)H(\mathcal{P}_i(t)). 
\end{align}

The simple flow rescaling enables efficient community detection at different Markov times with the search algorithm Infomap \cite{infomap}. While Infomap is designed to minimize the original map equation over possible network partitions, it can be applied to the reconstructed network that corresponds to the transition matrix for a given Markov time. This works for the continuous-time transition matrix $T_\mathrm{C}(t)$ in \cref{eq:continuoustransition} \cite{Schaub:2012ct}, as well as for its linearized form in \cref{eq:linearizedtransition}. While reconstructing the linearized transition matrix is much faster and does not densify the network, further improvement is possible. In fact, the reconstruction can be completely evaded. Since the self-links only indirectly affect the map equation for Markov time $t$ in \cref{eq:mapeqmarkovt} by reducing the transition rates between nodes and modules, exactly the same effect can be achieved by directly rescaling Infomap's internal representation of flows along links by a factor $t$. This is the approach we take. Infomap takes as input the original network and the Markov time $t$, calculates the ergodic node visit and transition rates, and then rescales the transition rates by a factor $t$ without any network reconstruction at all. 

\begin{figure}
\centering
\includegraphics[width=\columnwidth]{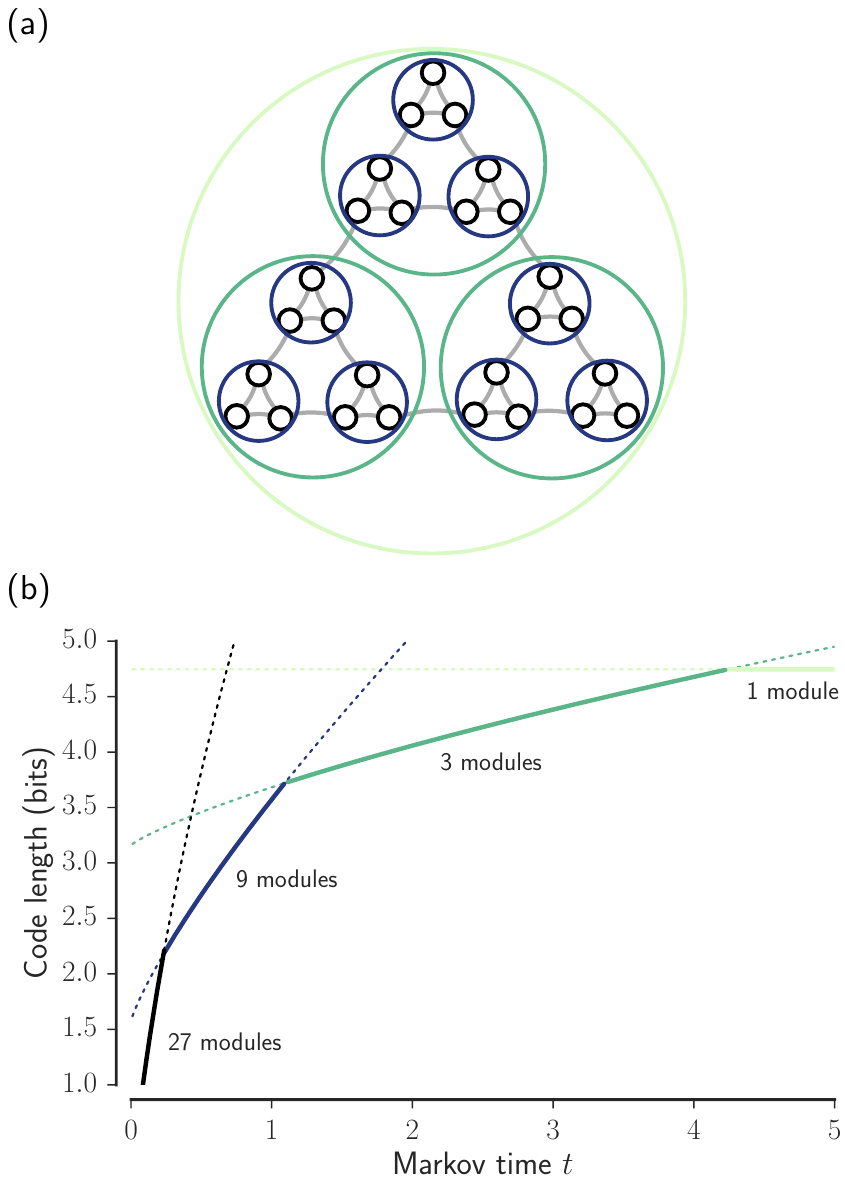}
\caption{(Color online) \label{fig:coderate}The Markov time sets the scale of the flow modules. (a) A schematic Sierpinski network with hierarchically nested modules. (b) The code length for different partitions indicated in the network as a function of the Markov time. The partition with the shortest code length for a given Markov time is highlighted.}
\end{figure}

Figure \ref{fig:coderate} shows an example with a Sierpinski network. For the shortest Markov times, putting every node in its own module gives the shortest code length. For longer Markov times, solutions with larger and larger modules give the shortest code length.  

\begin{figure}
\centering
\includegraphics[width=\columnwidth]{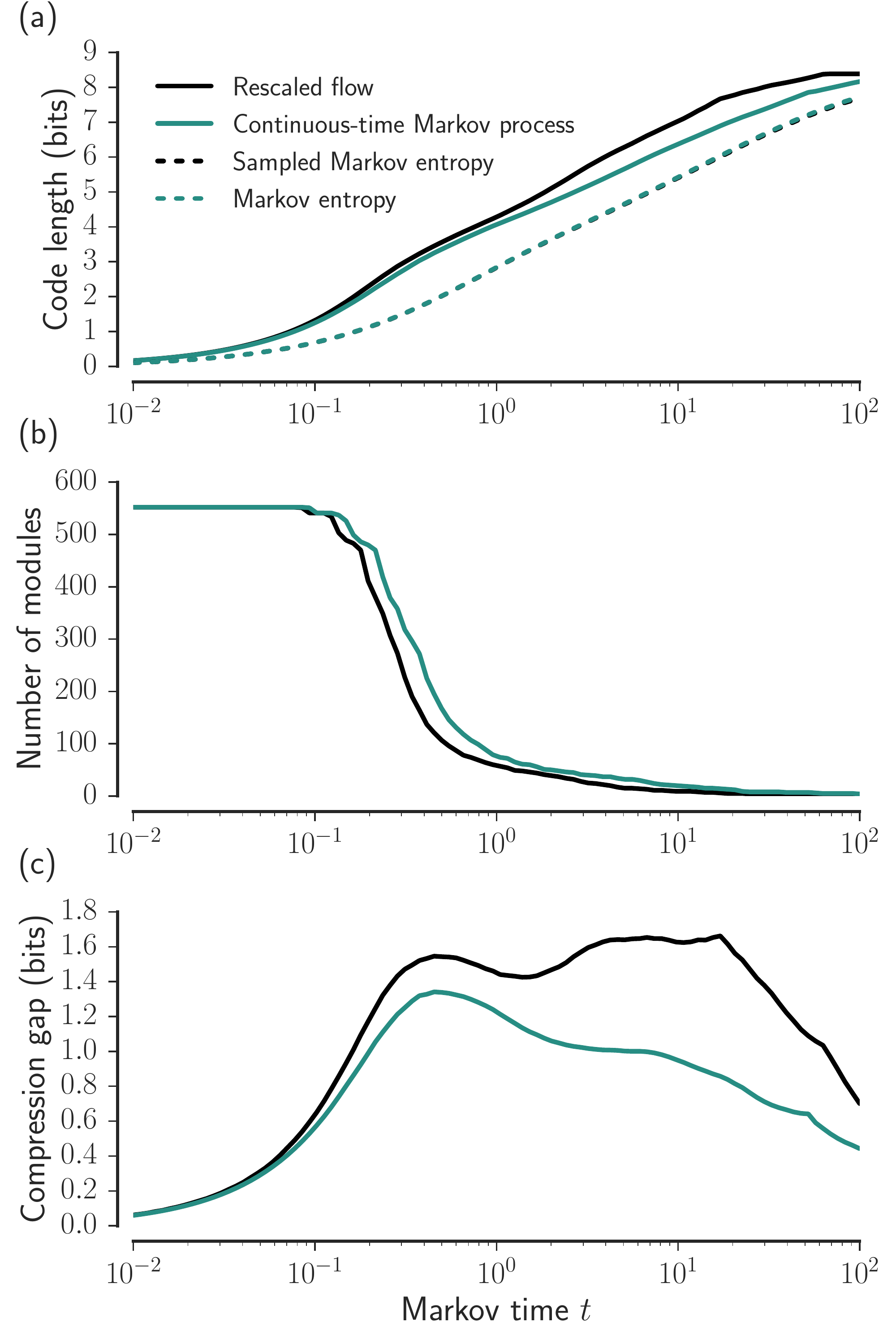}
\caption{(Color online) \label{fig:zoomlenscomp}Comparing flow rescaling with a continuous-time Markov process. Panels a-c show the effect on a weighted, undirected co-authorship network with 552 physicists \cite{esquivel2011compression}. Standard deviations are smaller than the line width.}
\end{figure}

The simple flow rescaling gives a slightly different encoding of the dynamics than the continuous-time Markov process \cite{Schaub:2012ct}. The flow rescaling only operates on transitions between nodes directly connected in the original network and does only indirectly consider transitions between nodes connected by multi-step trajectories. Contrarily, the continuous-time Markov process directly considers a spectrum of these trajectories. Their lengths are given by the transition matrix power in the expanded continuous-time transition matrix in \cref{eq:continuoustransition},
\begin{align}
T_\mathrm{C}(t) = e^{-t}I+te^{-t}T_\mathrm{D}+\frac{t^2e^{-t}}{2}T_\mathrm{D}^2 + \frac{t^3e^{-t}}{6}T_\mathrm{D}^3 + \cdots,
\end{align}
such that they are Poisson distributed with mean length $t$. From a coding perspective, the continuous-time transition matrix allows a random walker on a multi-step journey on the original network to move out of a module and back again between two encodings without triggering any module exit- and enter-codewords. In the flow rescaling approach, however, such moves will indeed be encoded. As a result, the continuous-time Markov process allows flows to stay longer within a given module and therefore typically gives smaller modules and shorter description length. Figure \ref{fig:zoomlenscomp} illustrates the effects of the different dynamics on a weighted, undirected co-authorship network with 552 physicists \cite{esquivel2011compression}. While the flow rescaling gives longer code lengths especially for longer Markov times (\cref{fig:zoomlenscomp}(a)), and somewhat larger modules for the same Markov time (\cref{fig:zoomlenscomp}(b)), the overall patterns are the same.

The network and problem at hand may set a natural Markov time, but often the most appropriate Markov time is unknown. Based on the rationale that good modular solutions should give good compressions, Schaub \textit{et al.\ } suggested to compare the code length of the modular description by the map equation at a given Markov time with the entropy rate of the corresponding Markov process,
\begin{align}
h_\mathrm{C}(t) = -\sum_{\alpha\beta}p_{\alpha}T_{\mathrm{C}\alpha\beta}(t)\log{T_{\mathrm{C}\alpha\beta}}(t),
\end{align}
which sets the lower limit on the description length \cite{Schaub:2012ct}. We use the same compression gap approach, but for better performance instead obtain the entropy rates at different Markov times by sampling random walks on the original network. That is, we repeatedly sample start nodes proportional to their ergodic visit-rates, and, for each start node, repeatedly perform random walks of lengths sampled from a Poisson distribution with expected length $t$. By averaging over the entropy of the final node for each start node, we can estimate the entropy rate of the continuous-time Markov process without constructing the corresponding continuous-time transition matrix $T_\mathrm{C}(t)$. Note that we can not use the linearized transition matrix in \cref{eq:linearizedtransition}, because the corresponding entropy rate is only a good estimate for $t < 1$ and does not converge to the entropy rate of the independent and identically distributed process for long Markov times, $\lim_{t \to \infty} h_\mathrm{C}(t) = -\sum_{\alpha}p_{\alpha}\log{p_{\alpha}} =  H(\mathcal{P})$, which is also the one-module solution of the map equation for any Markov time. Figure \ref{fig:zoomlenscomp}(a) shows that the sampled estimate performs well and practically overlaps with the Markov entropy obtained from the continuous-time transition matrix. Schaub \textit{et al.\ } looked at the relative compression gap \cite{Schaub:2012ct}, but to avoid inflating small differences for short Markov times, in \cref{fig:zoomlenscomp}(c) we show the absolute compression gap, $L(\mathsf{M},t)-h_\mathrm{C}(t)$. For this co-authorship network, the compression gaps indicate a local minimum just shorter than Markov time 2 for the rescaled flow and a local quasi-minimum just longer than Markov time 2 for the continuous-time Markov process. Interestingly, these Markov times correspond to about the same number of modules, since the flow rescaling generates slightly larger modules for the same Markov time.

Overall, the flow rescaling is in practice computationally much more efficient than the continuous-time Markov process, since the network must not be rebuilt for each Markov time. The continuous-time Markov process generates dense networks for long Markov times, which results in infeasible solutions for large networks. Contrarily, the flow rescaling has similar fast performance for all Markov times. However, for networks so sparse that random fluctuations can cause quenched modules \cite{lancichinetti2009community}, it can pay off to incorporate longer trajectories. Then extending the linearized transition matrix in \cref{eq:linearizedtransition} with quadratic terms from the continuous-time transition matrix $T_\mathrm{C}(t)$ in \cref{eq:continuoustransition} can provide an efficient compromise between the slower continuous-time Markov process, which makes the network denser, and the faster flow rescaling, which maintains the network density.

\subsection*{The map equation for bipartite networks}

A complete projection of a bipartite network with \emph{primary nodes} and \emph{feature nodes} into a unipartite network with only primary node gives an overload of links already for moderately dense networks \cite{alzahrani2014community}. Here we explore three ways to overcome this problem for the map equation framework: projecting by rescaling the Markov time, treating the network as unipartite, and projecting by sampling important links. 

\begin{figure}[htb]
\centering
\includegraphics[width=\columnwidth]{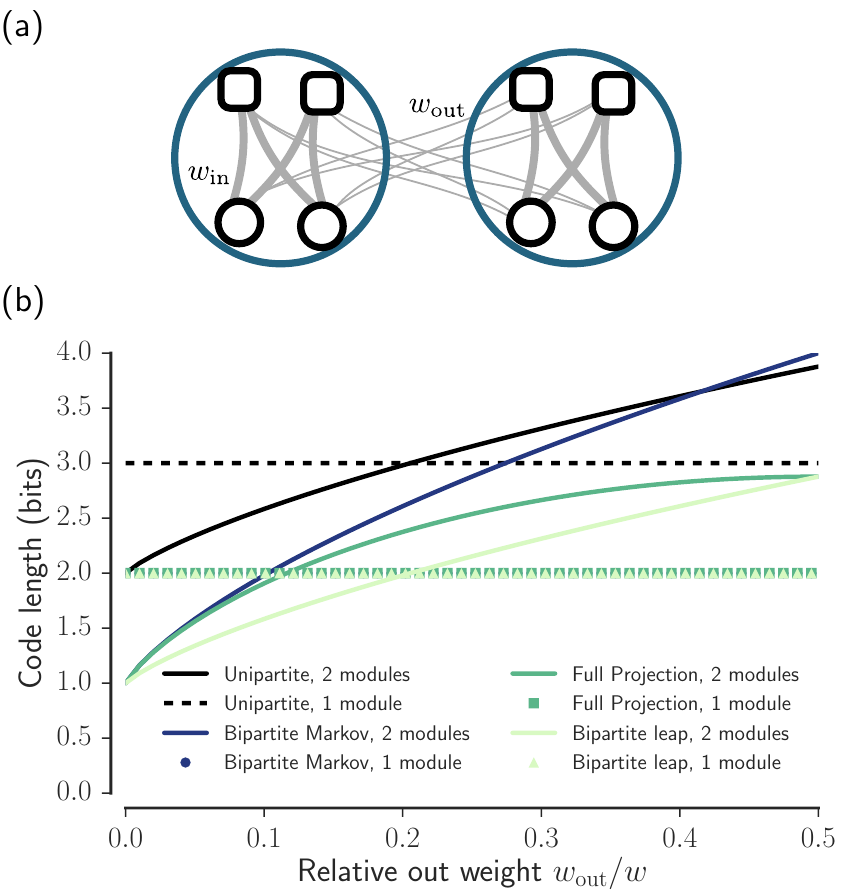}
\caption{(Color online) \label{fig:bipsweep}Projecting bipartite networks corresponds to doubling the Markov time and increases the scale of flow modules. (a) A schematic bipartite network with link weight $w_{\mathrm{in}}$ between primary nodes (circles) and feature nodes (squares) in the same community and  link weight $w_{\mathrm{out}}$ between nodes in different communities. (b) The code length for different bipartite dynamics and coding schemes as a function of the relative out weight.}
\end{figure}

Flow rescaling makes a projection effortless, because projecting a bipartite network into a unipartite network essentially corresponds to a rescaling of the Markov time. With Markov time 2, a random walker will take two steps between two encodings such that the exit and enter rates according to \cref{eq:markovexit,eq:markoventer} become
\begin{align}
q_{i\exit}(2) = 2q_{i\exit}\label{eq:bipexit},\\
q_{i\enter}(2) = 2q_{i\enter}\label{eq:bipenter}.
\end{align}
If such random walkers with a cycle of two are released on the primary nodes, only the primary node visits will be encoded. In this way, the map equation takes exactly the same form as in \cref{eq:mapeqmarkovt} with $t=2$, 
\begin{align}\label{eq:bipmapeq}
L(\mathsf{M},2) = q_{\enter}(2) H(\mathcal{Q}(2)) + \sum_{i=1}^{m}p_{i\intra}(2)H(\mathcal{P}_i(2)),
\end{align}
with the only difference that the visit rates of primary nodes double and the visit rates of feature nodes become 0. Therefore, with subscript $p$ for primary nodes and $f$ for features nodes,
\begin{align}
p_{\alpha,p}(2) &= 2 p_{\alpha}(1)\label{eq:bipmarkovp}\\
p_{\alpha,f}(2) &= 0.\label{eq:bipmarkovf}
\end{align}
Even if visits to feature nodes do not contribute to the code length, the flow rates between modules depend on their module assignments. Therefore, both primary nodes and feature nodes are clustered. Since the flow rescaling treats movements in and out of modules differently than with a projected and fully rebuilt network as described above, the projection with rescaled Markov time best approximates the full projection for small flows between modules (see \cref{fig:bipsweep}).

A similar approach to doubling the Markov time is to instead use random walkers that leap over every other node. That is, the dynamics take place on the full network with primary nodes and feature nodes as above, but only steps from feature nodes to primary nodes are accounted for. By rescaling the total visit rates to 1, the node visit rates take the same form as in \cref{eq:bipmarkovp,eq:bipmarkovf}, but the transition rates in \cref{eq:markovexit,eq:markoventer} now depend on the relative amount of flow that moves between modules from feature nodes to primary nodes. For undirected networks, the flow is equal in both directions such that the bipartite leap dynamics correspond to Markov time $t=1$ in \cref{eq:markovexit,eq:markoventer}. That is, the bipartite leap dynamics effectively correspond to the standard unipartite dynamics in which the node type is ignored as shown in \cref{fig:bipsweep}. While only encoding primary nodes offsets the code length compared to the unipartite dynamics, the compression gain between different modular solutions remains exactly the same for the schematic network in \cref{fig:bipsweep}. In general, the difference is so small that an approach based on the bipartite leap dynamics is superfluous, and we will instead use the unipartite dynamics when comparing different approaches.

The research question at hand will determine which approach should be favored. In the example in \cref{fig:bipsweep}, the two approaches that correspond to dynamics with Markov time 2, full projection and projection with rescaled Markov time, favor the two-module solution until about 10\% relative out-weight. Therefore, they can work well for sparse networks or interest in large-scale structures. Instead, the two approaches that correspond to Markov time 1, the unipartite and bipartite leap dynamics, favor the two-module solution until about 20\% relative out-weight. Therefore, they can work well for dense networks or interest in small-scale structures. 

With two methods that can work well at different scales, we now turn to a fast projection approach based on sampling of important links that resembles the method we used for estimating Markov entropies above. It is an adaptive method that can work well at a wider range of scales. Sampling of important links works well in practice, because most links in a weighted projection will carry redundant information for community detection. Therefore, only the important and non-redundant links must be sampled. Much like the Minhash approach \cite{broder2000min}, we seek to identify similar nodes of one type. In our case, nodes that are frequently visited in sequence by a random walker that performs two-step dynamics on a bipartite network. In detail, we associate each feature node with the top $X$ primary nodes selected by link weight, or randomly for ties as in unweighted networks. For each primary node, we take the top $X$ primary nodes associated with each of its connected feature node and include them in a candidate set. For each node in the candidate set, we compute the two-step random walk probability to go to other nodes also in the candidate set and create links to the top $Y$ nodes. For all experiments in this paper, we used $X=1,000$ and $Y=10$. For these choices, we found that the sampling approach can be both fast and accurate for dense as well as sparse networks.

\section*{Results and discussion}

To compare the three methods, we tested their performance on bipartite benchmark networks. To construct the bipartite benchmark networks, we built on the standard approach with a generative model for unipartite networks \cite{girvan2002community}. We assigned both \emph{primary nodes} and \emph{feature nodes} to communities and then added $k$ unweighted and undirected links between each primary node and $k_\mathrm{in}$ randomly chosen feature nodes in the same community and $k_\mathrm{out} = k - k_\mathrm{in}$ randomly chosen feature nodes in other communities. Specifically, we used 32 communities, each with 32 primary nodes with average degree 16, and varied the number of links between communities and the number of feature nodes for more or less sparse networks.

\begin{figure}[htb]
\centering
\includegraphics[width=\columnwidth]{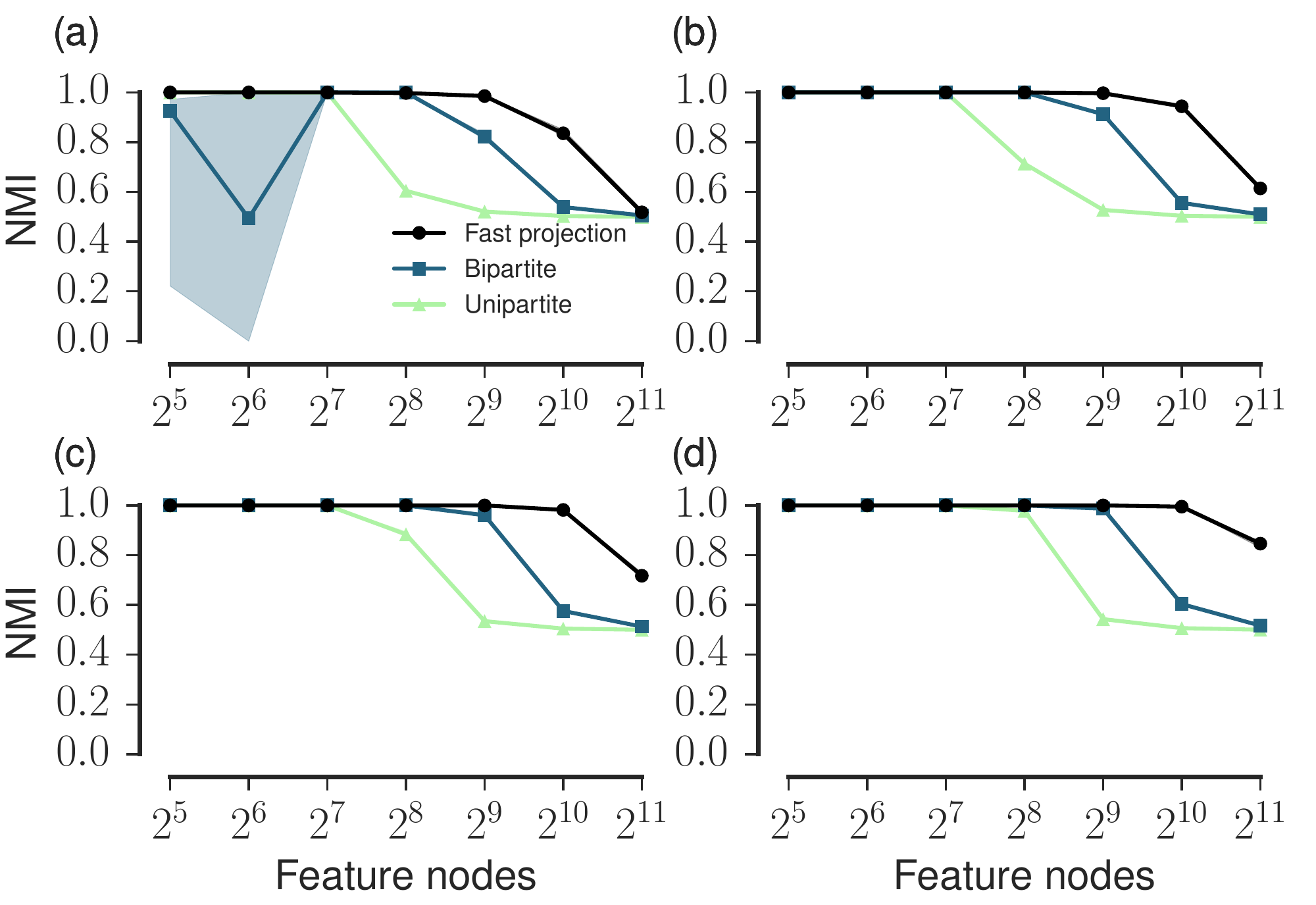}
\caption{(Color online) \label{fig:bipbench}Fast projection performs well on both sparse and dense bipartite benchmark networks. The performance of fast projection, the bipartite dynamics, and the unipartite dynamics measured by the normalized mutual information, NMI, as a function of the number of feature nodes and the number of links between communities, $k_\mathrm{in}=12$ in (a), 13 in (b), 14 in (c), and 15 in (d). Filled area represents standard deviation.}
\end{figure}

\begin{table*}[tb]
\centering
\caption{Comparing two-level and multilevel community detection of unipartite dynamics, bipartite dynamics, and fast projection applied to real-world bipartite networks. Modules for the multilevel solutions report the total number of modules across all levels. All result values are reported with two significant figures}
\label{table:realnets}
\begin{footnotesize}
\begin{tabular}{@{\extracolsep{\fill}}@{}lrrrr@{\hskip 1em}rrrr@{\hskip 1em}rrrr@{\hskip 1em}rrr@{}}
\mytoprule\noalign{\smallskip}
 & \multicolumn{3}{r}{\textbf{arXiv collaboration}} & & \multicolumn{3}{r}{\textbf{20 Newsgroups}} & & \multicolumn{3}{r}{\textbf{YouTube}} & & \multicolumn{3}{r}{\textbf{MovieLens}} \\ 
 Primary nodes & \multicolumn{3}{r}{16,726} && \multicolumn{3}{r}{17,856} && \multicolumn{3}{r}{94,238} && \multicolumn{3}{r}{6,040} \\
 Feature nodes & \multicolumn{3}{r}{22,015} && \multicolumn{3}{r}{78,198} && \multicolumn{3}{r}{30,087} && \multicolumn{3}{r}{3,900} \\
 Links & \multicolumn{3}{r}{58,595} && \multicolumn{3}{r}{1,873,331} && \multicolumn{3}{r}{293,360} && \multicolumn{3}{r}{1,000,209} \\ \noalign{\smallskip}
 & Unipart. & Bipart. & F. proj. && Unipart. & Bipart. & F. proj. && Unipart. & Bipart. & F. proj. & & Unipart. & Bipart. & F. proj. \\
 \cline{2-4}\cline{6-8}\cline{10-12}\cline{14-16}
 \textbf{Two-level} \\
 Modules & 3,100 & 2,200 & 2,500 && 740 & 36 & 660 && 9,500 & 7,900 & 7,100 && 250 & 1 & 35 \\ \noalign{\smallskip}
 \emph{NMI} \\
Unipartite & 1.00 &  &  && 1.00 &  &  && 1.00 &  &  &&  1.00 &  & \\
Bipartite & 0.91 & 1.00 & && 0.04 & 1.00 & && 0.59 & 1.00 & && 0.00 & 1.00 &  \\
Fast projection & 0.94 & 0.92 & 1.00 && 0.08 & 0.00 & 1.00 && 0.77 & 0.57 & 1.00 && 0.00 & 0.00 & 1.00 \\ \noalign{\smallskip} 
 \textbf{multilevel} \\
 Levels & 6 & 5 & 6 && 2 & 2 & 4 && 5 & 3 & 4 && 2 & 1 & 2 \\
 Modules & 7,300 & 3,100 & 4,200 && 740 & 36 & 900 && 12,000 & 8,000 & 8,000 && 250 & 1 & 35 \\ \noalign{\smallskip}
\emph{HNMI} \\
Unipartite & 1.00 & & && 1.00 &  &  && 1.00 &  &  && 1.00 &  &  \\
Bipartite & 0.66 & 1.00 & && 0.04 & 1.00 & && 0.25 & 1.00 & && 0.00 & 1.00 & \\
Fast projection & 0.66 & 0.58 & 1.00 && 0.02 & 0.00 & 1.00 && 0.59 & 0.23 & 1.00 && 0.00 & 0.00 & 1.00 \\ \mybottomrule
\end{tabular}
\end{footnotesize}
\end{table*}

The bipartite benchmark test reveals the effect of different effective Markov times (\cref{fig:bipbench}). Standard unipartite dynamics or the bipartite leap dynamics, which correspond to Markov time 1, work well down to relatively high number of links between communities as long as the number of feature nodes is limited. With increasing number of feature nodes, the network becomes sparser, and the dynamics generate quenched modules. The bipartite dynamics, which approximates a projection of the network and corresponds to Markov time 2, cannot resolve communities as accurately as the unipartite approach for dense networks with high number of links between communities (\cref{fig:bipbench}). On the other hand, the bipartite dynamics can better handle sparse networks with many feature nodes. Finally, fast projection effectively adapts the Markov time and handles both dense and sparse networks on par or better than the approaches with fixed Markov times. Unless the research question calls for a specific Markov time, fast projection stands out as a good choice.

Finally we applied the three different methods on four real-world bipartite networks (see \cref{table:realnets}). For each network we report the number of primary and feature nodes and the number of links. We applied both two-level and multilevel community detection with the search algorithm Infomap \cite{infomap}. In the first approach, we forced Infomap to find two-level solutions, while in the second approach we let Infomap find the multilevel solution with the optimal number of nested levels for best compression of the dynamics. We report the standard NMI for the two-level approach \cite{danon2005comparing} and the generalized NMI for the multilevel approach \cite{perotti2015hierarchical}. For the multilevel approach, we also report the number of levels for the best solution as well as the total number of modules across all levels. The real bipartite networks include an author-paper network, \textbf{arXiv collaboration} \cite{newman2001structure}, a document-word network, \textbf{20 Newsgroups} \cite{20news}, a user-group network, \textbf{YouTube} \cite{youtube}, and a user-movie network, \textbf{MovieLens} \cite{herlocker1999algorithmic}. All networks are popular for performing benchmark experiments. 

The comparison between the methods applied on real networks confirms the results from the synthetic benchmark tests: unipartite dynamics reveal more and smaller modules than bipartite dynamics because of the inherently shorter Markov time of unipartite dynamics (\cref{table:realnets}). Again, fast projection effectively adapts its Markov time and the network determines whether fast projection most resembles unipartite or bipartite dynamics. For the 20 Newsgroups and MovieLens networks, the NMI scores are low because the solutions of the unipartite and bipartite dynamics basically have one dominating module and many tiny modules. The two-level results carry over to the multilevel solutions, and unipartite dynamics typically give deeper solutions than bipartite dynamics. Overall, fast projection adapts the effective Markov time and can handle both sparser and denser networks. 

\section*{Conclusions}

We introduced an efficient method to perform community detection of network flows at different Markov times. The method takes advantage of the information-theoretic machinery of the map equation and handles projections of bipartite networks as well. In synthetic and real-world networks, we showed how modifying the Markov times influences the size of the identified communities. Depending on the network and question at hand, a shorter Markov time with smaller communities in deeper multilevel structures or longer Markov time with larger communities in shallower multilevel structures may be more appropriate. For bipartite networks, we also introduced a fast projection approach that effectively adapts the Markov time for robust communities. While current methods require expensive and often infeasible network reconstructions, the introduced methods offer efficient alternatives applicable to large networks.

We have made the code available in the Infomap software package, which also includes efficient community detection for varying Markov times of higher-order Markov processes \cite{infomap}.

\begin{acknowledgments}
M.R.\ was supported by the Swedish Research Council grant 2012-3729. We are grateful to Carl Bergstrom and Renaud Lambiotte for helpful discussions.
\end{acknowledgments}



%

\end{document}